\documentclass[12pt]{article}
\usepackage{amstex}

\columnwidth\textwidth

\begin{document}

\newcommand{\be}{\begin{equation}}
\newcommand{\ee}{\end{equation}}
\newcommand{\beann}{\begin{eqnarray*}}
\newcommand{\eeann}{\end{eqnarray*}}
\newcommand{\bea}{\begin{eqnarray}}
\newcommand{\eea}{\end{eqnarray}}
\newcommand{\nn}{\nonumber}
\newcommand{\ben}{\begin{enumerate}}
\newcommand{\een}{\end{enumerate}}
\newtheorem{df}{Definition}
\newtheorem{thm}{Theorem}
\newtheorem{lem}{Lemma}
\newtheorem{prop}{Proposition}
\begin{titlepage}

\noindent
\hspace*{11cm} BUTP-01/25 \\
\vspace*{1cm}
\begin{center}
{\LARGE Pair of null gravitating shells II. \newline
Canonical theory and embedding variables} 

\vspace{2cm}

P. H\'{a}j\'{\i}\v{c}ek and I.~Kouletsis \\
Institute for Theoretical Physics \\
University of Bern \\
Sidlerstrasse 5, CH-3012 Bern, Switzerland \\
\vspace*{2cm}

December 2001 \\ \vspace*{1cm}

\nopagebreak[4]

\begin{abstract}
  The study of the two shell system started in our first paper ``Pair of null
  gravitating shells I'' is continued. An action functional for a single shell
  due to Louko, Whiting and Friedman is generalized to give appropriate
  equations of motion for two and, in fact, any number of spherically
  symmetric null shells, including the cases when the shells intersect. In
  order to find the symplectic structure for the space of solutions described
  in paper I, the pull back to the constraint surface of the Liouville form
  determined by the action is transformed into new variables. They consist of
  Dirac observables, embeddings and embedding momenta (the so-called
  Kucha\v{r} decomposition). The calculation includes the integration of a set
  of coupled partial differential equations. A general method of solving the
  equations is worked out.
\end{abstract}

\end{center}

\end{titlepage}

\section{Introduction}
\label{sec:intro}
The present paper is the second in a series dedicated to the two-shell
system. The first paper, Ref.~\cite{I}, will be referred to as I
henceforth. The general motivation and aims are explained in I. Some solutions
to the classical equations of motion containing two shells have been studied
in I with two main results: First, certain coordinates have been chosen on
the space of all solutions; they are candidates for a complete set of
Dirac observables of the system. Second, all symmetries of the system that are
associated with diffeomorphisms between solution spacetimes have been
found. These symmetries will define the most interesting observables as well
as the true Hamiltonian of the system.

To proceed, the space of solutions ought to be promoted to a physical phase
space, which is a symplectic manifold. Without knowledge of the symplectic
structure, not even the decision can be made whether some physical coordinates
are differentiable. Therefore, we turn in this paper to the canonical theory.

As far as we know, no action functional for the two-shell system exists in
literature. For a single spherically symmetric null shell, a Hamilton action
principle has been studied by Kraus and Wilczek \cite{K-W} and by Louko,
Whiting and Friedman \cite{L-W-F}. We shall start from this action and
generalize it for two and more shells. The generalization is easy, and a proof
is given that the resulting equations of motion admit the correct solutions
for the system, including all solutions\footnote{These solutions are a subset
  of the solutions admitted by the equations of motion due to the additional
  assumption of the regular center on the left.} described in I.  This, of
course, is not a full justification for the choice of action. As it is well
known, given dynamical equations admit many non-equivalent action principles
from which they follow.  An example for a single massive shell is investigated
in \cite{Hm}. A direct reduction of the Einstein-Hilbert plus a shell-matter
action to spherical symmetry might lead to a non-equivalent theory.
Ref.~\cite{jez} is an example of such an action principle without symmetry,
but it is valid only for a single shell.  The generalization of the
Louko-Whiting-Friedman action to more shells and intersections is simpler than
that of the action in \cite{jez} so we have chosen to start from it.  We
assume that both actions lead to equivalent theories.

The plan of the paper is as follows. In Sec.~\ref{sec:canon}, we propose our
generalization of the Louko-Whiting-Friedman action principle to include two
and, in fact, any number of in-going and out-going shells. We prove that the
variation of this action gives the equations of motion that we need.

Sec.~\ref{sec:Liouville} begins with the calculation of Poisson brackets
between the parameters that have been chosen as coordinates on the space of
solutions in I. The strategy is to pull back the Liouville form defined by our
action to the constraint surface, and to transform variables so that the new
ones consist of the parameters, embeddings and embedding momenta. This is the
so-called Kucha\v{r} decomposition studied in \cite{H-Kij}. A similar
transformation has been accomplished in \cite{H-Kie} for the case of a single
shell.  The calculation includes integration of a set of partial differential
equations of the first order. The integration method used in \cite{H-Kie}
seems to be rather closely associated with the particular problem studied and
the gauge chosen there. We shall find a general method that works for any
number of shells and in any double-null gauge.

\section{Canonical formalism}
\label{sec:canon}
\subsection{The action}
\label{sec:action}
The action functional for a single light-like shell and its gravitational
field has been written down in Ref.~\cite{L-W-F}. It is modified below so that
it describes two such shells:
\begin{equation}
  S_2 = \int d\tau\left[{\mathbf p}_1\dot{\mathbf r}_1 + {\mathbf
  p}_2\dot{\mathbf r}_2 + \int_0^\infty d\rho(P_\Lambda\dot{\Lambda} +
  P_R\dot{R} - H_2)\right]\ .
\label{81.1}
\end{equation}
The Hamiltonian has the same overall form as in \cite{L-W-F} 
\[
  H_2 = {\mathcal NH} + {\mathcal N}^\rho{\mathcal H}_\rho + {\mathcal
  N}_\infty E_\infty\ ,
\]
but the new constraints are
\begin{eqnarray}
  {\mathcal H} & = & \frac{\Lambda P_\Lambda^2}{2R^2} - \frac{P_\Lambda
  P_R}{R} + 
  \frac{RR''}{\Lambda} - \frac{RR'\Lambda'}{\Lambda^2} + \frac{R^{\prime
  2}}{2\Lambda} - \frac{\Lambda}{2} \nn \\
  && + \frac{\eta_1{\mathbf p}_1}{\Lambda}\delta(\rho - {\mathbf r}_1) +
  \frac{\eta_2{\mathbf p}_2}{\Lambda}\delta(\rho - {\mathbf r}_2)\ ,
\label{81.2} \\
  {\mathcal H}_\rho & = & P_RR' - P'_\Lambda\Lambda - {\mathbf p}_1\delta(\rho
  - {\mathbf r}_1) - {\mathbf p}_2\delta(\rho - {\mathbf r}_2)\ .
\label{81.3}
\end{eqnarray}
The momenta of the shells are ${\mathbf p}_1$ and ${\mathbf p}_2$, their
radial coordinates are ${\mathbf r}_1$ and ${\mathbf r}_2$ and their radial
directions are $\eta_1$ and $\eta_2$. The dot denotes the derivative with
respect to $\tau$ and the prime that with respect to $\rho$. The ``volume''
variables $\Lambda$, $R$, $P_\Lambda$, $P_R$, $\mathcal N$ and ${\mathcal
  N}^\rho$ are the same as in \cite{L-W-F} and \cite{H-Kie}. The meaning of
the variables $\Lambda$, $R$, $\mathcal N$ and ${\mathcal N}^\rho$ can be
inferred from the spacetime metric
\begin{equation}
  ds^2 = -{\mathcal N}^2d\tau^2 + \Lambda^2(d\rho + {\mathcal N}^\rho d\tau)^2
  + R^2d\Omega^2.
\label{L2.1}
\end{equation}
The momenta conjugate to the configuration variables $\Lambda$ and $R$ are
\begin{equation}
  P_\Lambda = -\frac{R}{\mathcal N}(\dot{R} - {\mathcal N}^\rho R')\ ,
\label{L2.5b}
\end{equation}
and
\begin{equation}
  P_R = -\frac{\Lambda}{\mathcal N}(\dot{R} - {\mathcal N}^\rho R') -
  \frac{R}{\mathcal N}[\dot{\Lambda} - ({\mathcal N}^\rho\Lambda)']\ .
\label{L2.5c}
\end{equation}

Compared to ref.~\cite{L-W-F}, the only change is the presence of one more
term in the Liouville form and of a corresponding term in each constraint.
Each of these terms has the same form as in \cite{L-W-F}, but now, they carry
indices so that the two shells can be distinguished.

It is important to state the differentiability conditions on the shells,
especially at a possible shell crossing (if $\eta_1 \neq \eta_2$). One can
assume as in \cite{L-W-F} that the gravitational variables are smooth
functions of $\rho$, with the exception that ${\mathcal N}'$, $({\mathcal
  N}^\rho)'$, $\Lambda'$, $R'$, $P_\Lambda$ and $P_R$ may have finite
discontinuities at isolated values of $\rho$. The coordinate loci of the
discontinuities are smooth functions of $\tau$ for each shell. This follows
from (i) the conditions at isolated single shell points which are the same as
in \cite{L-W-F}, (ii) from the conditions at any shell crossing as described
in Sec.~2.2 of I and (iii) from the corresponding choice of foliation:
the metric with respect to coordinates $\tau$ and $\rho$ may be piecewise
smooth and everywhere continuous.

\subsection{Equations of motion}
\label{sec:eqn}
The rest of the section will be devoted to a check that the variation of the
action (\ref{81.1}) yields the proper dynamical equations for two shells.
These equations have to imply (a) that the geometry outside the shells is the
Schwarzschild one and (b) that the shells move along null surfaces. These two
properties, together with the continuity of the metric everywhere and the
assumption of a regular center on the left, have been used in Sec.~2 of I to
construct all solutions. The continuity of the metric is part of the
definition of the configuration space, and the existence of a left regular
center is an assumption of our model, but the two other properties have to
result from the variation of the action.

The continuity of the metric defines a $C^1$ class of coordinates; we shall
work with such coordinates in this section.  We also assume that the foliation
coordinates $\tau$ and $\rho$ are $C^1$.  The variations of the action
(\ref{81.1}) with respect to the gravitational variables $\mathcal N$,
${\mathcal N}^\rho$, $R$, $\Lambda$, $P_R$ and $P_\Lambda$ give the
constraints,
\begin{equation}
  {\mathcal H} = 0\ ,\quad {\mathcal H}_\rho = 0
\label{L2.8a}
\end{equation}
and the dynamical equations
\begin{eqnarray}
  \dot{\Lambda} & = & {\mathcal N}\left(\frac{\Lambda P_\Lambda}{R^2} -
  \frac{P_R}{R}\right) + ({\mathcal N}^\rho\Lambda)'\ ,
\label{L2.9a} \\
  \dot{R} & = & -\frac{{\mathcal N}P_\Lambda}{R} + {\mathcal N}^\rho R'\ ,
\label{L2.9b} \\
  \dot{P}_\Lambda & = & \frac{\mathcal N}{2}\left[-\frac{P_\Lambda^2}{R^2} -
  \left(\frac{R'}{\Lambda}\right)^2 + 1 + \frac{2\eta_1{\mathbf
  p}_1}{\Lambda^2}\delta(\rho - {\mathbf r}_1) + \frac{2\eta_2{\mathbf
  p}_2}{\Lambda^2}\delta(\rho - {\mathbf r}_2)\right]  \nn \\
  &&- \frac{{\mathcal
  N}'RR'}{\Lambda^2} + {\mathcal N}^\rho P_\Lambda'\ ,
\label{L2.9c} \\
  \dot{P}_R & = & {\mathcal N}\left[\frac{\Lambda P_\Lambda^2}{R^3} -
  \frac{P_\Lambda P_R}{R^2} - \left(\frac{R'}{\Lambda}\right)'\right] -
  \left(\frac{{\mathcal N}'R}{\Lambda}\right)' + ({\mathcal N}^\rho P_R)'\ .
\label{L2.9d}
\end{eqnarray}

Outside the shells ($\rho \neq {\mathbf r}_1, \rho \neq {\mathbf r}_2$), they
coincide with the equations of \cite{L-W-F} and imply that the geometry there
is the Schwarzschild one corresponding to some, as yet arbitrary, value of
Schwarzschild mass parameter.

\subsubsection{Outside the crossing}
Let us first consider a Cauchy surface that does not contain a crossing point
of the two shells, ${\mathbf r}_1 \neq {\mathbf r}_2$. This includes data for
crossing as well as for parallel shells.

The equations that we obtain at $\rho = {\mathbf r}_i$, $i = 1,2$, result
partly from setting the coefficients at the delta functions $\delta(\rho -
{\mathbf r}_1)$ and $\delta(\rho - {\mathbf r}_2)$ in
Eqs.~(\ref{L2.8a})--(\ref{L2.9d}) equal to zero, and partly from the
variations of the action (\ref{81.1}) with respect to ${\mathbf r}_i$ and
${\mathbf p}_i$.  If we use the notation $\delta_{\mathbf r}(X)$ for the
coefficient at the $\delta(\rho - {\mathbf r})$ in the expression $X$, then
the equations read:
\begin{eqnarray}
  0 & = & \delta_{{\mathbf r}_i}\left(\frac{RR''}{\Lambda}\right) +
  \frac{\eta_i{\mathbf p}_i}{\Lambda_{\rho={\mathbf r}_i}}\ ,
\label{2a.1} \\
  0 & = & \delta_{{\mathbf r}_i}(\Lambda P_\Lambda') + {\mathbf p}_i\ ,
\label{2a.2} \\
  \delta_{{\mathbf r}_i}(\dot{P}_\Lambda) & = & \left(\frac{\mathcal
  N}{\Lambda^2}\right)_{\rho={\mathbf r}_i} \eta_i{\mathbf p}_i +({\mathcal
  N})_{\rho={\mathbf r}_i} \delta_{{\mathbf r}_i}(P_\Lambda')\ ,
\label{2a.3} \\
  \delta_{{\mathbf r}_i}(\dot{P}_R) & = & -({\mathcal N})_{\rho={\mathbf r}_i}
  \delta_{{\mathbf r}_i}\left(\left[\frac{R'}{\Lambda}\right]'\right) -
  \delta_{{\mathbf r}_i}\left(\left[\frac{{\mathcal
  N}'R}{\Lambda}\right]'\right) + \delta_{{\mathbf r}_i}([{\mathcal N}^\rho
  P_R]')\ ,
\label{2a.4} \\
  \dot{\mathbf r}_i & = & \eta_i\left(\frac{\mathcal
  N}{\Lambda}\right)_{\rho={\mathbf r}_i} - ({\mathcal N}^\rho)_{\rho={\mathbf
  r}_i}\ ,
\label{2a.5} \\
  \dot{\mathbf p}_i & = & {\mathbf p}_i\left({\mathcal N}^\rho -
  \frac{\eta_i{\mathcal N}}{\Lambda}\right)_{\rho={\mathbf r}_i}\ .
\label{2a.6} 
\end{eqnarray}
By inspection, the equations with any fixed value of $i$ coincide with the
one-shell equations of \cite{L-W-F}. Thus, outside the shells, as well as at
the shells outside the crossing, we obtain equations that have been shown in
\cite{L-W-F} to imply the required properties (a) and (b).

To obtain expressions for the momenta ${\mathbf p}_i$ conjugate to ${\mathbf
  r}_i$, we can use the constraints (\ref{2a.1}) and (\ref{2a.2}) because we
shall always work only at the constraint surface within the present series of
papers. We obtain the equations
\begin{eqnarray}
  R({\mathbf r}_i)\Delta_{{\mathbf r}_i}(R') + \eta_i{\mathbf p}_i & = & 0\ ,
\label{84a.1} \\
  \Lambda({\mathbf r}_i)\Delta_{{\mathbf r}_i}(P_\Lambda) + {\mathbf p}_i & =
  & 0\ .
\label{84a.2}
\end{eqnarray}
Here, we use the notation $\Delta_x(X)$ for the jump $\lim_{\rho\rightarrow
  x+}X - \lim_{\rho\rightarrow x-}X$ at the shell point $\rho = x$.

In any double-null coordinates $U$ and $V$, the metric has the form
\begin{equation}
  ds^2 = -AdUdV + R^2d\Omega^2
\label{metric}
\end{equation}
and the transformation formulae are:
\begin{eqnarray}
  \Lambda & = & \sqrt{-AU'V'}\ , 
\label{1*} \\
  {\mathcal N} & = & -\frac{\dot{U}V'-\dot{V}U'}{2U'V'}\sqrt{-AU'V'}\ , 
\label{2*} \\
  {\mathcal N}^\rho & = & \frac{\dot{U}V'+\dot{V}U'}{2U'V'}\ .
\label{3*}
\end{eqnarray}
It follows that
\begin{eqnarray}
  R' & = & R_{,U}U' + R_{,V}V'\ ,
\label{84a.3}  \\
  P_\Lambda & = & \frac{R}{\Lambda}(R_{,U}U' - R_{,V}V')\ .
\label{84a.4} 
\end{eqnarray}
The last equation is derived from Eq.~(\ref{L2.5b}) and
(\ref{1*})--(\ref{3*}).  We shall often need the following simple Lemma.
\begin{lem}
  Let $\varphi$ be a continuous function of $U$ and $V$ and let its
  derivatives have jumps only at $U = U({\mathbf r})$ and $V = V({\mathbf
    r})$, being themselves continuous elsewhere. Then $\varphi_{,U}$ is a
  continuous function of $V$ everywhere except for $U = U({\mathbf r})$ where
  the jump $\Delta(\varphi_{,U})$ is a continuous function of $V$ for all $V$
  (even at $V = V({\mathbf r})$). Similarly for $\varphi_{,V}$.
\end{lem}
The Lemma follows immediately from the continuity of $\varphi$.

Applying the Lemma to the function $R$, which is continuous, we find that
$\Delta(R_{,U}) \neq 0, \Delta(R_{,V}) = 0$ for $\eta_i = +1$ and
$\Delta(R_{,U}) = 0, \Delta(R_{,V}) \neq 0$ for $\eta_i = -1$.

Substituting this into Eqs.~(\ref{84a.1}) and (\ref{84a.2}), one finds the
desired expressions:
\begin{equation}
  {\mathbf p}_i = - R({\mathbf r}_i)\Delta_{{\mathbf r}_i}(R_{,U})U'
\label{84a.5}
\end{equation}
for $\eta_i = +1$ and 
\begin{equation}
  {\mathbf p}_i = R({\mathbf r}_i)\Delta_{{\mathbf r}_i}(R_{,V})V'
\label{84a.6}
\end{equation}
for $\eta_i = -1$. The orientation of the functions $U$, $V$, $\tau$ and
$\rho$ is such that $U' < 0$ and $V' > 0$ (see \cite{H-Kie}).

\subsubsection{At the crossing: the constraints}
Let us now turn to the Cauchy surfaces that contain a crossing point of the
shells, where we have ${\mathbf r}_1 = {\mathbf r}_2 = {\mathbf r}$. Here,
there is only one $\delta$-function, $\delta(\rho - {\mathbf r})$, and there
is only one coefficient in all variation equations at $\delta(\rho - {\mathbf
  r})$. Thus, instead of two equations, we obtain only one in each case. A new
analysis is in order.

Let us first consider the constraints (\ref{L2.8a}). The
coefficients of $\delta(\rho - {\mathbf r})$ in both constraints are
\begin{eqnarray*}
  \delta_{\mathbf r}({\mathcal H}) & = & \frac{\eta_1{\mathbf
  p}_1}{\Lambda({\mathbf r})} + \frac{\eta_2{\mathbf
  p}_2}{\Lambda({\mathbf r})} + \frac{R({\mathbf r})}{\Lambda({\mathbf
  r})}\Delta_{\mathbf r}(R')\ , \\
  \delta_{\mathbf r}({\mathcal H}^\rho) & = & -{\mathbf p}_1 - {\mathbf p}_2 -
  \Lambda({\mathbf r})\Delta_{\mathbf r}(P_\Lambda)\ .
\end{eqnarray*}

To study the equations $\delta_{\mathbf r}({\mathcal H}) = 0$ and
$\delta_{\mathbf r}({\mathcal H}^\rho) = 0$, it is advantageous to choose some
double-null coordinates in which the metric is continuous. Similar tactic has
been pursued in the Appendix B of \cite{L-W-F}. The expressions (\ref{84a.3})
and (\ref{84a.4}) for $R'$ and $P_\Lambda$ in terms of general double-null
coordinates yield now
\begin{eqnarray*}
  \Delta_{\mathbf r}(R') & = & \Delta_{\mathbf r}(R_{,U})U'({\mathbf r}) +
  \Delta_{\mathbf r}(R_{,V})V'({\mathbf r})\ , \\
  \Delta_{\mathbf r}(P_\Lambda) & = & \frac{R({\mathbf r})}{\Lambda({\mathbf
  r})} \left[\Delta_{\mathbf r}(R_{,U})U'({\mathbf r}) -
  \Delta_{\mathbf r}(R_{,V})V'({\mathbf r})\right]\ .
\end{eqnarray*}
Thus, the equation $\delta_{\mathbf r}({\mathcal H}) = 0$ becomes
\begin{equation}
  R({\mathbf r})\Delta_{\mathbf r}(R_{,U})U'({\mathbf r}) + R({\mathbf r})
  \Delta_{\mathbf r}(R_{,V})V'({\mathbf r}) = -\eta_1{\mathbf p}_1 -
  \eta_2{\mathbf p}_2 
\label{87.1}
\end{equation}
and $\delta_{\mathbf r}({\mathcal H}^\rho) = 0$ is
\begin{equation}
  R({\mathbf r})\Delta_{\mathbf r}(R_{,U})U'({\mathbf r}) - R({\mathbf r})
  \Delta_{\mathbf r}(R_{,V})V'({\mathbf r}) = -{\mathbf p}_1 -
  {\mathbf p}_2\ .
\label{87.2}
\end{equation}
Let $\eta_1 = -\eta_2 = 1$. Then, we obtain from Eqs.~(\ref{87.1}) and
(\ref{87.2}) 
\begin{eqnarray*}
  {\mathbf p}_1 & = & -R({\mathbf r})\Delta_{\mathbf r}(R_{,U})U'({\mathbf
  r})\ , 
  \\ 
  {\mathbf p}_2 & = & R({\mathbf r})\Delta_{\mathbf r}(R_{,V})V'({\mathbf r})\
  . 
\end{eqnarray*}
If $\eta_1 = -\eta_2 = -1$, then
\begin{eqnarray*}
  {\mathbf p}_1 & = & R({\mathbf r})\Delta_{\mathbf r}(R_{,V})V'({\mathbf r})\
  , 
  \\ 
  {\mathbf p}_2 & = & -R({\mathbf r})\Delta_{\mathbf r}(R_{,U})U'({\mathbf
  r})\ . 
\end{eqnarray*}
This can be summarized by
\begin{eqnarray}
  {\mathbf p}_{\text{out}} & = & -R({\mathbf r})\Delta_{\mathbf
  r}(R_{,U})U'({\mathbf r})\ , 
\label{88.1}  \\ 
  {\mathbf p}_{\text{in}} & = & R({\mathbf r})\Delta_{\mathbf
  r}(R_{,V})V'({\mathbf r})\ . 
\label{88.2}
\end{eqnarray}
Notice that these equations are exactly the same as Eqs.~(\ref{84a.5}) 
and (\ref{84a.6}) for shells outside the crossing. They
can serve as definition of the momenta everywhere.
From Lemma 1 and Eqs.~(\ref{88.1}) and (\ref{88.2}) we can infer that
${\mathbf p}_{\text{out}}$ is continuous along $U = U({\mathbf r})$ and
${\mathbf p}_{\text{in}}$ along $V = V({\mathbf r})$.

The fact that we obtain only two relations from the four equations
$\delta_{{\mathbf r}_1}({\mathcal H}) = 0$, $\delta_{{\mathbf r}_2}({\mathcal
H}) = 0$, $\delta_{{\mathbf r}_1}({\mathcal H}_\rho) = 0$ and $\delta_{{\mathbf
r}_2}({\mathcal H}_\rho) = 0$ outside the crossing is due to the mutual
dependence of ${\mathcal H}$- and ${\mathcal H}_\rho$-equations (see
\cite{L-W-F}) in this case. At the crossing the ${\mathcal H}$- and ${\mathcal
H}_\rho$-equations are independent.

\subsubsection{At the crossing: the shell variables}
Next, we study the variation of the action (\ref{81.1}) with respect to the
shell variables ${\mathbf r}_i$ and ${\mathbf p}_i$. The variation is to be
done so that ${\mathbf r}_1$ and ${\mathbf r}_2$ are considered as independent
variables even where ${\mathbf r}_1 = {\mathbf r}_2$; therefore, one must
vary first and only then set ${\mathbf r}_1 = {\mathbf r}_2$. The resulting
Euler-Lagrange equations then are
\begin{equation}
  \dot{\mathbf r}_i = \frac{{\mathcal N}({\mathbf r})}{\Lambda({\mathbf r})}
  \eta_i - {\mathcal N}^\rho({\mathbf r})
\label{89.1}
\end{equation}
and 
\begin{equation}
  \dot{\mathbf p}_i = -\left(\frac{{\mathcal N}(\rho)}{\Lambda(\rho)}\eta_i -
  {\mathcal N}^\rho(\rho)\right)'_{\rho={\mathbf r}}{\mathbf p}_i\ . 
\label{89.2}
\end{equation}
Eq.~(\ref{89.1}) is analogous to Eq.~(\ref{2a.5}) and Eq.~(\ref{89.2}) to
Eq.~(\ref{2a.6}).

Let us assume that the index $i$ is chosen in such a way that $\eta_i$ remains
constant along the trajectory of the $i$-th shell for each $i = 1,2$. This is
not the only possible assumption if the shells cross each other. For example,
one could choose $i=1$ for the innermost shell and $i=2$ for the outermost
one. Then the subsequent argument has to be modified.

The interpretation of Eq.~(\ref{89.1}) is the same as in \cite{L-W-F}, if we
use our convention on the index $i$. The right-hand side of Eq.~(\ref{89.1})
is then continuous along each shell because the functions ${\mathcal N}$,
${\mathcal N}^\rho$, and $\Lambda$ are components of the metric in the
coordinates $\tau$ and $\rho$. The vector tangential to the shell,
\[
  l^\mu_i = (1,\dot{\mathbf r}_i,0,0),
\]
is a null vector as a consequence of Eq.~(\ref{89.1}). Hence, the equation
completes the dynamical information that is necessary for construction of
solutions. The remaining equations for the shell have, therefore, to follow
from Einstein's equations outside the shells, Eqs.~ (\ref{88.1}), (\ref{88.2})
and (\ref{89.1}).

Let us extend the definition of the quantity $l^\rho_i$ to a whole
neighborhood of $i$-th shell by
\[
  l^\rho_i(\rho) := \frac{{\mathcal N}(\rho)}{\Lambda(\rho)}\eta_i -
  {\mathcal N}^\rho(\rho)\ .
\]
The function $l^\rho_i(\rho)$ can be calculated in terms of the metric and
embeddings for an arbitrary double-null gauge. Using the transformation
Eqs.~(\ref{1*})--(\ref{3*}), we obtain
\begin{equation}
  l^\rho_i(\rho) = -\frac{\dot{U}}{U'}  
\label{91.1}
\end{equation}
for $\eta_i = +1$ and 
\begin{equation}
  l^\rho_i(\rho) = -\frac{\dot{V}}{V'}  
\label{91.2}
\end{equation}
for $\eta_i = -1$. At the shells, this follows also directly from the
relations $U(\tau,{\mathbf r}(\tau))=$ const for $\eta_i = +1$ and
$V(\tau,{\mathbf r}(\tau))=$ const for $\eta_i = -1$, and it is analogous to
Eq.~(B3) in \cite{L-W-F}. The right-hand sides of Eqs.~(\ref{91.1}) and
(\ref{91.2}) are, of course, continuous functions in a neighborhood of $i$-th
shell because we have required the foliation to be $C^1$.

Let us turn to Eq.~(\ref{89.2}). As in \cite{L-W-F}, the first question is
whether the right-hand side is continuous across the shells; if it is not we
have a bad ambiguity. Let $\eta_i = +1$ and let us consider the question of
derivatives of the function $l^\rho_i(\rho)$ with respect to $\rho$ from both
sides of $i$-th shell (the argument is completely analogous to that in
Appendix B of \cite{L-W-F}):
\[
  \left(\frac{\partial l^\rho_i}{\partial \rho}\right)_{\rho={\mathbf r}\pm}
  = -\frac{1}{U'}\left(\dot{U}' - \frac{\dot{U}}{U'}U''\right)_{\rho={\mathbf
  r}\pm}\ .
\]
We have, however,
\begin{eqnarray*}
  \left(\dot{U}' - \frac{\dot{U}}{U'}U''\right)_{\rho={\mathbf
  r}\pm} & = & (\dot{U}')_{\rho={\mathbf r}\pm} + \dot{\mathbf
  r}(U'')_{\rho={\mathbf r}\pm} \\
  & = & \frac{d}{d\tau}U'(\tau,{\mathbf r}(\tau))\ ,
\end{eqnarray*}
and this is, of course, well defined. Hence, Eq.~(\ref{89.2}) is not
ambiguous. The proof for $\eta_i =-1$ is similar.

Next, we show that Eq.~(\ref{89.2}) is satisfied if Eqs.~(\ref{88.1}) and
(\ref{88.2}) hold and the function $R$ is continuous at the crossing. As we
have seen, Lemma 1 then implies that the jumps $\Delta(R_{,U})$ and
$\Delta(R_{,V})$ are continuous along the corresponding shells, even through
the crossing point. Let $\eta_i=+1$ and let us substitute Eq.~(\ref{84a.5})
for ${\mathbf p}_i$ into Eq.~(\ref{89.2}). A simple calculation gives
\[
  \frac{d}{d\tau}[\Delta_{\mathbf r}((R^2)_{,U})] U'(\tau,{\mathbf r}(\tau)) =
  0 
\]
along the shell and also at the crossing point. However, outside the crossing,
this equation follows from other equations (see \cite{L-W-F}), so it does not
contain any new information, even at the crossing: if $\Delta_{\mathbf
r}((R^2)_{,U})$ is time independent from both sides and continuous everywhere,
then it must be also constant everywhere.

\subsubsection{At the crossing: $\Lambda$ and $R$}
The remaining dynamical equations for the shells are obtained by varying the
action (\ref{81.1}) with respect to the variables $\Lambda$ and $R$ and
setting the coefficients at $\delta(\rho - {\mathbf r})$ in the resulting
equations equal to zero. The variations with respect to $P_\Lambda$ and $P_R$
give just definitions of $P_\Lambda$ and $P_R$, which do not contain
$\delta(\rho - {\mathbf r})$. Thus, at the crossing point, we obtain the
equations:
\begin{equation}
  \delta_{\mathbf r}(\dot{P}_\Lambda) - {\mathcal N}^\rho({\mathbf
  r})\delta_{\mathbf r}(P'_\Lambda) = \frac{{\mathcal N}({\mathbf
  r})}{\Lambda^2({\mathbf r})} (\eta_1{\mathbf p}_1 + \eta_2{\mathbf p}_2)\ ,
\label{94.1}
\end{equation}
and 
\begin{equation}
  \delta_{\mathbf r}(\dot{P}_R) - {\mathcal N}^\rho({\mathbf
  r})\delta_{\mathbf r}(P'_R) + \frac{{\mathcal N}({\mathbf
  r})}{\Lambda({\mathbf r})}\delta_{\mathbf r}(R'') + \frac{R({\mathbf
  r})}{\Lambda({\mathbf r})}\delta_{\mathbf r}({\mathcal N}'') = 0\ .
\label{94.2}
\end{equation}
while there are four equations outside the crossing, (\ref{2a.3}) and
(\ref{2a.4}), that can be cast as follows:
\begin{equation}
  \delta_{{\mathbf r}_i}(\dot{P}_\Lambda) - {\mathcal N}^\rho({\mathbf
  r}_i)\delta_{{\mathbf r}_i}(P'_\Lambda) = \frac{{\mathcal N}({\mathbf
  r}_i)}{\Lambda^2({\mathbf r}_i)} \eta_i{\mathbf p}_i\ ,
\label{94.3}
\end{equation}
and 
\begin{equation}
  \delta_{{\mathbf r}_i}(\dot{P}_R) - {\mathcal N}^\rho({\mathbf
  r}_i)\delta_{{\mathbf r}_i}(P'_R) + \frac{{\mathcal N}({\mathbf
  r}_i)}{\Lambda({\mathbf r}_i)}\delta_{{\mathbf r}_i}(R'') + \frac{R({\mathbf
  r}_i)}{\Lambda({\mathbf r}_i)}\delta_{{\mathbf r}_i}({\mathcal N}'') = 0
\label{94.4}
\end{equation}
for each $i = 1,2$.

Let us study the four discontinuous functions $R'$, ${\mathcal N}'$,
$P_\Lambda$ and $P_R$, the $\tau$ and $\rho$ derivatives of which feature
under the $\delta_{\mathbf r}$ signs in Eqs.~(\ref{94.1})--(\ref{94.4}). It is
advantageous to express them as functions of $U$ and $V$ and of the foliation
$U(\tau,\rho)$ and $V(\tau,\rho)$. For $R'$ and $P_\Lambda$,
Eqs.~(\ref{84a.3}) and (\ref{84a.4}) yield the desired expressions. For
${\mathcal N}'$, we have immediately
\begin{equation}
  {\mathcal N}' = {\mathcal N}_{,U}U' + {\mathcal N}_{,V}V'\ .
\label{95.2}
\end{equation}
For $P_R$, Eq.~(\ref{L2.5c}) and (\ref{1*})--(\ref{3*}) yield
\[
  P_R = R_{,U}U' - R_{,V}V' + \frac{RA_{,U}}{2A}U' - \frac{RA_{,V}}{2A}V' +
 \frac{RU''}{2U'} - \frac{RV''}{2V'}\ .
\]
We observe that
\[
  \frac{U''}{U'} - \frac{V''}{V'} =
  \left[\ln\left(\frac{-U'}{V'}\right)\right]' =
  \left[\ln\left(\frac{-U'}{V'}\right)\right]_{,U}U' +
  \left[\ln\left(\frac{-U'}{V'}\right)\right]_{,V}V'\ , 
\]
where $\ln(-U'/V)$ is considered as a (continuous) function of $U$
and $V$. Thus, we obtain, finally:
\begin{multline}
  P_R = \frac{1}{2AR}\left(R^2A\right)_{,U}U' -
  \frac{1}{2AR}\left(R^2A\right)_{,V}V' \\
  +\frac{R}{2}\left[\ln\left(\frac{-U'}{V'}\right)\right]_{,U}U' + \frac{R}{2}
  \left[\ln\left(\frac{-U'}{V'}\right)\right]_{,V}V'\ .
\label{95.3}
\end{multline}
Eqs.~(\ref{84a.3}), (\ref{84a.4}), (\ref{95.2}) and (\ref{95.3}) show that
each discontinuous function is a sum of terms of the form $\varphi\psi_{,U}$ or
$\varphi\psi_{,V}$, where $\varphi$ and $\psi$ are continuous functions of $U$
and $V$. The jump structure of such terms is given by the Lemma 1.

Let us denote by $R'_{\text{out}}$, ${\mathcal N}'_{\text{out}}$,
$P_{\Lambda\text{out}}$ and $P_{R\text{out}}$ the sum of all terms in the
right-hand sides of Eqs.~(\ref{84a.3}), (\ref{84a.4}), (\ref{95.2}) and
(\ref{95.3}) that contain only the $U$-derivatives, and similarly for
$R'_{\text{in}}$ etc. Thus, e.g., 
\[
  P_{R\text{out}} = \frac{1}{2AR}\left(R^2A\right)_{,U}U' +
  \frac{R}{2}\left[\ln\left(\frac{-U'}{V'}\right)\right]_{,U}U'\ , 
\]
etc. Eqs.~(\ref{94.1}) and (\ref{94.2}) can then be written as follows:
\begin{multline}
  \left[\delta_{\mathbf r}(({P}_{\Lambda\text{out}})\spdot) - {\mathcal
  N}^\rho({\mathbf r})\delta_{\mathbf r}((P_{\Lambda\text{out}})')\right] + 
  \left[\delta_{\mathbf r}(({P}_{\Lambda\text{in}})\spdot) - {\mathcal
  N}^\rho({\mathbf r})\delta_{\mathbf r}((P_{\Lambda\text{in}})')\right] \\
  = \frac{{\mathcal N}({\mathbf
  r})}{\Lambda^2({\mathbf r})} (\eta_{\text{out}}{\mathbf p}_{\text{out}} +
  \eta_{\text{in}}{\mathbf p}_{\text{in}})\ , 
\label{97.1}
\end{multline}
and 
\begin{multline}
  \delta_{\mathbf r}(({P}_{R\text{out}})\spdot) - {\mathcal N}^\rho({\mathbf
  r})\delta_{\mathbf r}((P_{R\text{out}})')) + \frac{{\mathcal N}({\mathbf
  r})}{\Lambda({\mathbf r})}\delta_{\mathbf r}((R'_{\text{out}})') +
  \frac{R({\mathbf 
  r})}{\Lambda({\mathbf r})}\delta_{\mathbf r}(({\mathcal N}'_{\text{out}})')
  \\ 
  +\delta_{\mathbf r}(({P}_{R\text{in}})\spdot) - {\mathcal N}^\rho({\mathbf
  r})\delta_{\mathbf r}((P_{R\text{in}})') + \frac{{\mathcal N}({\mathbf
  r})}{\Lambda({\mathbf r})}\delta_{\mathbf r}((R'_{\text{in}})') +
  \frac{R({\mathbf 
  r})}{\Lambda({\mathbf r})}\delta_{\mathbf r}(({\mathcal N}'_{\text{in}})')
  = 0\ ,  
\label{97.2}
\end{multline}
while Eqs.~(\ref{94.3}) and (\ref{94.4}) become, for the out-going shell,
\begin{equation}
  \delta_{{\mathbf r}_{\text{out}}}(({P}_{\Lambda\text{out}})\spdot) -
  {\mathcal N}^\rho({\mathbf 
  r}_{\text{out}})\delta_{{\mathbf r}_{\text{out}}}((P_{\Lambda 
  \text{out}})') = \frac{{\mathcal N}({\mathbf 
  r}_{\text{out}})}{\Lambda^2({\mathbf r}_{\text{out}})}
  \eta_{\text{out}}{\mathbf p}_{\text{out}}\ ,   
\label{97.3}
\end{equation}
and
\begin{multline}
  \delta_{{\mathbf r}_{\text{out}}}(({P}_{R\text{out}})\spdot) - {\mathcal
  N}^\rho({\mathbf 
  r}_{\text{out}})\delta_{{\mathbf r}_{\text{out}}}((P_{R\text{out}})') \\
  + \frac{{\mathcal N}({\mathbf 
  r}_{\text{out}})}{\Lambda({\mathbf r}_{\text{out}})}\delta_{{\mathbf
  r}_{\text{out}}}((R'_{\text{out}})') + \frac{R({\mathbf 
  r}_{\text{out}})}{\Lambda({\mathbf r}_{\text{out}})}\delta_{{\mathbf
  r}_{\text{out}}}(({\mathcal N}'_{\text{out}})') = 0   
\label{97.4}
\end{multline}
and similarly for the in-going shell. The reason is that only the derivatives
of the out-terms give contributions to $\delta$'s along the out-going shell
because the in-terms are continuous, according to Lemma 1.

The left-hand sides of Eqs.~(\ref{97.3}) and (\ref{97.4}) are obtained from
the jumps of $U$ derivatives of continuous functions in a way that is
continuous along the out-going shell. The jumps themselves are continuous
along the shell because of Lemma 1. However, the out-parts of
Eqs.~(\ref{97.1}) and (\ref{97.2}) are made in the same way from the jumps of
$U$ derivatives at ${\mathbf r}_{\text{out}} = {\mathbf r}$. Because of the
continuity of all terms along the shell, the out-part of the left-hand side of
Eq.~(\ref{97.1}) is the ${\mathbf r}_{\text{out}} \rightarrow {\mathbf r}$
limit of the left-hand side of Eq.~(\ref{97.3}). Analogous claims hold for
left-hand sides of Eqs.~(\ref{97.2}) and (\ref{97.4}) as well as for the
in-terms. Moreover, because of the continuity of ${\mathbf p}_{\text{out}}$
and $\eta_{\text{out}}$ along the out-going shell, the out-term on the
right-hand side of Eq.~(\ref{97.1}) is the limit of the right-hand side of
Eq.~(\ref{97.3}).

It follows that each of Eqs.~(\ref{97.1}) and (\ref{97.2}) can be considered
as sum of two equations, one being the limit ${\mathbf r}_{\text{out}}
\rightarrow {\mathbf r}$ of the corresponding out-going shell
Eqs.~(\ref{97.3}) and (\ref{97.4}), the other being the same limit of an
in-going shell equation. Hence, if the out- and in-going shell equations hold
for all values of ${\mathbf r}_{\text{out}}$ and ${\mathbf r}_{\text{in}}$
outside the crossing point, then Eqs.~(\ref{97.1}) and (\ref{97.2}) are also
valid. This is implied by the continuity conditions on the phase-space
variables.

However, Eqs.~(\ref{97.3}) and (\ref{97.4}) are satisfied because they follow
from Eqs.~(\ref{94.3}) and (\ref{94.4}), and these, in turn, follow from other
dynamical equations. This has been shown in \cite{L-W-F} for a single shell
and the proof is, formally, the same in our case because all one-shell
equations outside the shell crossing are formally identical with equations in
\cite{L-W-F}.

To summarize: We have shown that the action (\ref{81.1}) gives proper
dynamical equations for the two-shell system.

\section{The Liouville form at the constraint surface}
\label{sec:Liouville}
Our final aim is to calculate the Poisson brackets between Dirac observables
such as $M_m$, $M_r$, $v_{m2}-v_{m1}$ and $v_r$ defined in I. Our method will
employ the property of the pull-back ${\Theta}_{\Gamma}$ of the Liouville form
${\Theta}$ to the constraint surface $\Gamma$ that it depends only on the
Dirac observables.  Its external differential then defines the symplectic form
of the physical phase space.  In the present section, we develop some general
tools in this line.

The Liouville form of the action (\ref{81.1}) can be written as follows:
\begin{equation}
  \Theta = {\mathbf p}_1\dot{\mathbf r}_1 + {\mathbf p}_2\dot{\mathbf r}_2 -
  {\mathcal N}_\infty E_\infty + \int_0^\infty d\rho\,(P_\Lambda\dot{\Lambda}
  + P_R\dot{R})  \, . 
\label{4,1}
\end{equation}
We have included the boundary part of the Hamiltonian
into $\Theta$; the form of this part justifies such inclusion (cf.~\cite{K},
\cite{H-Kie}). We can now start to transform (\ref{4,1}) into the Kucha\v{r}
variables corresponding to arbitrary double-null gauge.
While $\Theta_\Gamma$ does not depend of the gauge and dependent degrees of
freedom, the old variables contain them and we must use the complete
transformation; it goes from the variables $R$, $\Lambda$, $P_R$, $P_\Lambda$
to observables, gauge variables and dependent variables. Still, the resulting
$\Theta_\Gamma$ contains then only the former and none of the latter.

To write down such a transformation, we shall use a particular Kucha\v{r}
decomposition; for definition and existence, see \cite{H-Kij}. To this aim, we
shall choose an arbitrary double-null gauge, represented by the coordinates
$U$ and $V$. We also let the choice of Dirac observables open, denoting them
by $o^k$, $k=1,2,\dots,2N$. The complete set of final variables consists,
therefore, of the physical variables $o^k$, the gauge variables represented by
the embeddings $(U(\rho),V(\rho))$ and the dependent variables represented by
the embedding momenta $P_U(\rho)$ and $P_V(\rho)$; at the constraint surface,
$P_U(\rho) = P_V(\rho) = 0$.

To start the calculation, we just need to know that the metric (\ref{metric})
depends on $o^i$:
\[
  A = A(U,V;o)\ ,\quad R = R(U,V;o)\ .
\]
Then we use the transformation formulae (\ref{1*})--(\ref{3*}) and the
definitions (\ref{L2.5b}) and (\ref{L2.5c}) of the momenta. Such calculation
has been already carried out in \cite{H-Kie} so we just take over the relevant
general formulae from there.

The calculation in \cite{H-Kie} then leaves this general stage and proceeds by
making a particular choice of $o$'s as well as of the gauge $U$ and $V$. It
grows rather complicated and can be accomplished by some miraculous tricks
whose nature seems to be closely connected with the particular choices of 
gauge and observables made. The main purpose of the present section is to 
reveal a general structure
that underlies the tricks and that is entirely general.

\subsection{The volume part}
The form (\ref{4,1}) can be divided into a boundary part (the first three
terms on the right-hand side) and the volume parts. Each volume part is
associated with a particular component of the space between the shells; it has
the form
\[
  \int_a^b d\rho\,(P_\Lambda\dot{\Lambda} + P_R\dot{R})\ ,
\]
where $a$ and $b$ are values of the coordinate $\rho$ at the boundary of the
volume. For example, $a = 0$ and $b = {\mathbf r}_1$, or $a = {\mathbf r}_2$,
$b = \infty$, etc.

The only ``volume variables'' among the final set are the embedding ones,
$U(\rho)$ and $V(\rho)$, and the embedding momenta $P_U(\rho)$ and
$P_V(\rho)$. However, $\Theta_\Gamma$ cannot contain $U(\rho)$ and $V(\rho)$
because they are gauge variables; still less can it contain $P_U(\rho)$ and
$P_V(\rho)$ because they vanish at $\Gamma$. Hence, we expect that the volume
parts of $\Theta_\Gamma$ can all be reduced to some boundary terms. A general
account of such a reduction is now given.

We can see that the form of the volume parts is independent of the system,
more precisely, of the number of the shells in the system. Hence, we can use
the methods of Ref.~\cite{H-Kie} for its transformation to the new variables.

For any double-null gauge, the equations
\begin{eqnarray}
  4RR_{,UV} + 4R_{,U}R_{,V} + A & = & 0\ ,
\label{5,1}  \\
  AR_{,UU} - A_{,U}R_{,U} & = & 0\ ,
\label{5,2} \\
  AR_{,VV} - A_{,V}R_{,V} & = & 0
\label{5,3}
\end{eqnarray}
represent the condition that the transformation is performed at the constraint
surface $\Gamma$ (cf.~\cite{H-Kie}, Eqs.~(32)--(34)). 

For the transformation of any volume part into the new variables, we can make
the ansatz
\begin{equation}
  \Theta_a^b\vert_\Gamma = \int_a^b d\rho\,[(f\dot{U} + g\dot{V} +
  h_i\dot{o}^i)' + \dot{\varphi}]\ ,
\label{ans}
\end{equation}
as in \cite{H-Kie}. In the expressions for the functions $f$, $g$, $h_i$ and
$\varphi$, we can separate the terms with $\rho$-derivatives of the embeddings
$U(\rho),V(\rho)$ from terms in which $U(\rho)$ and $V(\rho)$ are not
differentiated. The form of the transformation between the variables
$\Lambda$, $R$, $P_\Lambda$ and $P_R$ on one side and $U$, $V$, $P_{U}$,
$P_{V}$ and $o^i$ on the other then imply that
 \begin{eqnarray}
f &=& \frac{RR_{,U}}{2}\ln\left(-\frac{U'}{V'}\right)
 +F(U,V,o^i)\ ,
\label{6,1} \\
g &=& \frac{RR_{,V}}{2}\ln\left(-\frac{U'}{V'}\right)+G(U,V,o^i)\ ,
\label{6,2}\\
h_i &=& \frac{RR_{,i}}{2}\ln\left(-\frac{U'}{V'}\right)
 +H_i(U,V,o^i)\ ,
\label{6,3} \\
\varphi &=& RR_{,U}U'-RR_{,V}V'-\frac{R}{2}(R_{,U}U'+R_{,V}V')
 \ln\left(-\frac{U'}{V'}\right)
\nonumber
\\
& & \; -FU'-GV'+\phi(U,V,o^i)\ ,
\label{6,4}
\end{eqnarray}
and (see Eqs.~(45)--(48) of \cite{H-Kie}) 
\begin{eqnarray}
F_{,V}-G_{,U} &=& \frac{R}{2A}(2AR_{,UV}-A_{,U}R_{,V}-A_{,V}R_{,U})\ ,
\label{6,5} \\ 
H_{i,U}-F_{,i} &=& -\frac{R}{2A}(2AR_{,iU}-A_{,i}R_{,U}-A_{,U}R_{,i})\
,\label{6,6}\\ 
H_{i,V}-G_{,i} &=& \frac{R}{2A}(2AR_{,iV}-A_{,i}R_{,V}-A_{,V}R_{,i})\ ,
\label{6,7}\\ 
\phi &=& 0\ . \label{6,8}
\end{eqnarray}
Eqs.~(\ref{6,5})--(\ref{6,8}) are valid for any double-null gauge.  

The ansatz (\ref{ans}) leads to the following transformation of the volume
part:
\[
  \Theta^b_a\vert_\Gamma = (f\dot{U} + g\dot{V} + h_i\dot{o}^i -
  \varphi\dot{b})_{\rho=b} - (f\dot{U} + g\dot{V} + h_i\dot{o}^i -
  \varphi\dot{a})_{\rho=a} + \frac{d}{d\tau}\left(\int_a^b
  d\rho\,\varphi\right)\ .
\]
By ignoring the total time derivative we are thus left with an equivalent form
that contains only boundary terms:
\begin{equation}
  \Theta^b_a\vert_\Gamma = (f\dot{U} + g\dot{V} + h_i\dot{o}^i -
  \varphi\dot{b})_{\rho=b} - (f\dot{U} + g\dot{V} + h_i\dot{o}^i -
  \varphi\dot{a})_{\rho=a}\ .
\label{bound}
\end{equation}
The boundary values $a$ and $b$ of $\rho$ depend, in general, on the time
parameter $\tau$ as they can also describe positions of shells.  If the
intermediate boundaries are chosen to coincide with the positions of shells,
the total Liouville form $\Theta$ reduces to the sum of contributions from (i)
$\rho = {\bf r}_i$ (corresponding to points $a$, $b$ at which the embedding
$U(\rho), V(\rho)$ intersects a shell), (ii) $\rho = 0$ (corresponding to the
center $R=0$) and (iii) $\rho \rightarrow \infty$ (corresponding to asymptotic
infinity).  The boundary between two adjacent spacetime regions ${\cal M}_K$
and ${\cal M}_{K+1}$ is either a light-like hypersurface (defined by an
in-going or an out-going shell) or a crossing point (defined by the
intersection of two shells of different $\eta$).

In this way, $\Theta_\Gamma$ can be transformed to a sum of boundary terms,
provided that a solution to the system of differential equations
(\ref{6,5})--(\ref{6,7}) can be found. Let us study these equations.

\subsection{Properties of functions $F$, $G$ and $H_i$}
Let us first establish some general properties of the functions $F$, $G$ and
$H_i$ that follow from Eqs.~(\ref{6,5})--(\ref{6,7}).

\subsubsection{Freedom in the functions $F$, $G$, $H_i$}
Eqs.~(\ref{6,5})--(\ref{6,7}) form an inhomogeneous system of linear partial
differential equations of first order. Eqs.~(\ref{5,1})--(\ref{5,3})
constitute the integrability condition for the system. The general solution of
the system can be written as a sum of a particular solution and a general
solution of the homogeneous system. Let $F^0$, $G^0$ and $H^0_i$ be a
particular solution, and let $F^1$, $G^1$ and $H^1_i$ be a solution of the
homogeneous equations:
\begin{eqnarray}
  F^1_{,V} - G^1_{,U} & = 0\ ,
\label{101,1} \\
  H^1_{i,U} - F^1_{,i} & = 0\ ,
\label{101,2} \\
  H^1_{i,V} - G^1_{,i} & = 0\ .
\label{101,3}
\end{eqnarray}
Let us define the function $c_{ij}$ by
\begin{equation}
  c_{ij} := H^1_{i,j} - H^1_{j,i}
\label{101,4}
\end{equation}
and study its properties. Immediately from the definition, we have
\begin{equation}
  c_{ij} = -c_{ji}\ ,
\label{101,5}
\end{equation}
and 
\begin{equation}
  c_{ij,k} + c_{jk,i} + c_{ki,j} = 0\ .
\label{101,6}
\end{equation}
The derivative of Eq.~(\ref{101,2}) with respect to $o^j$ with subsequent
anti-symmetrization in the indices $i$ and $j$ yield
\[
  c_{ij,U} = 0\ .
\]
Using Eq.~(\ref{101,3}) in a similar way gives
\[
  c_{ij,V} = 0\ .
\]
Hence, $c_{ij}$ depends only on Dirac's observables. Eqs.~(\ref{101,5}) and
(\ref{101,6}) imply then that there is a function $C_i(o)$ such
that 
\[
  c_{ij} = C_{i,j} - C_{j,i}\ .
\]

Let us choose an arbitrary $C_i(o)$. Then the functions $F^1$, $G^1$ and
$H^1_i$ have to satisfy the equations
\begin{eqnarray}
  F^1_{,V} - G^1_{,U} & = &0\ ,
\label{102,1} \\
  H^1_{i,U} - F^1_{,i} & = &0\ ,
\label{102,2} \\
  H^1_{i,V} - G^1_{,i} & = &0\ ,
\label{102,3} \\
  H^1_{i,j} - H^1_{j,i} & = &C_{i,j} - C_{j,i}\ ,
\label{102,4}
\end{eqnarray}
Again, this is an inhomogeneous linear differential system. The following is
clearly a particular solution:
\[
  F^1 = G^1 = 0,\quad H^1_i = C_i\ .
\]
Any solution of the corresponding homogeneous system has, however, the form 
\[
  F^1 =  W_{,U},\quad G^1 = W_{,V},\quad H^1_i = W_{,i}\ ,
\]
where $W$ is an arbitrary function of $U$, $V$, and $o^i$.

We have shown: Let $F^0$, $G^0$ and $H^0_i$ be a solution to the system
(\ref{6,5})--(\ref{6,7}). Then any other solution $F$, $G$, $H_i$ has the
form
\begin{equation}
  F = F^0 + W_{,U}\ ,\quad G = G^0 + W_{,V}\ ,\quad H_i = H^0_i + W_{,i} +
  C_i\ ,
\label{102,5}
\end{equation}
where $W$ is an arbitrary function of the variables $U$, $V$ and $o^i$, and
$C_i$ is an arbitrary function of Dirac's observables. 

If we substitute the solution (\ref{102,5}) into Eq.~(\ref{ans}) for
$\Theta^b_a\vert_\Gamma$, we obtain for the terms containing the functions $W$
and $C_i$ (the sum of these terms is denoted by
$\delta\Theta^b_a\vert_\Gamma$):
\[
  \delta\Theta^b_a\vert_\Gamma = \int_a^b d\rho\,[(W_{,U}\dot{U} +
  W_{,V}\dot{V} + W_{,i}\dot{o^i} + C_i\dot{o}^i)' + (-W_{,U}U' -
  W_{,V}V')\spdot]\ .
\]
However, Dirac's observables do not depend on $\rho$, so $(C_i\dot{o}^i)' = 0$
and $W_{,U}U' - W_{,V}V' = w'$, where we have defined 
\[
  w(\rho,\tau) := W(U(\rho,\tau),V(\rho,\tau),o^i(\tau))\ .
\]
Using this, we have
\[
  \delta\Theta^b_a\vert_\Gamma = \int_a^b d\rho\,[(\dot{w})' - (w')\spdot] =
  0
\]
because all functions are $C^\infty$ in the space between the shells. Hence,
each solution (\ref{102,5}) leads to the same Liouville form.

The $W$- and $C_i$-part of the first parenthesis in Eq.~(\ref{bound}) is,
however, a total time derivative:
\begin{multline*}
  (W_{,U}\dot{U} + W_{,V}\dot{V} + W_{,i}\dot{o^i} + C_i\dot{o}^i +
  W_{,U}U'\dot{b} - W_{,V}V'\dot{b})_{\rho=b} = \\
  \frac{\partial w(b,\tau)}{\partial b}\dot{b} + \frac{\partial
  w(b,\tau)}{\partial \tau} = \frac{d}{d\tau}w(b(\tau),\tau)\ ,
\end{multline*}
which, in general, is non zero.  Similar result holds for the second
parenthesis. Hence, different solutions lead to equivalent boundary Liouville
forms.

\subsubsection{Gauge transformation of $F$, $G$ and $H_i$}
Two different double-null gauges lead to two different transformation of a
given volume part of the Liouville form. We can, therefore, ask how the
functions $F$, $G$ and $H_i$ are transformed if the gauge changes. The leading 
idea of course is that the Liouville form itself does not change.

A general gauge transformation between two double-null gauges, $\tilde{U}$,
$\tilde{V}$ and $U$, $V$, reads
\begin{equation}
  U = X(\tilde{U},o)\ ,\quad V = Y(\tilde{V},o)\ ,
\label{7,1}
\end{equation}
where $X$ and $Y$ are suitable functions; the inverse transformation can be
written as
\begin{equation}
  \tilde{U} = \tilde{X}(U,o)\ ,\quad \tilde{V} = \tilde{Y}(V,o)\ .
\label{8,1}
\end{equation}

The form $f\dot{U} + g\dot{V} + h_i\dot{o}^i$ transforms under the change
(\ref{8,1}) of variables as follows
\[
  f\dot{U} + g\dot{V} + h_i\dot{o}^i = \tilde{f}\dot{\tilde{U}} +
  \tilde{g}\dot{\tilde{V}} + \tilde{h}_i\dot{o}^i\ ,
\]
where
\[
  f = \tilde{f}\tilde{X}_{,U}\ , \quad g = \tilde{g}\tilde{Y}_{,V}\ ,
\]
and 
\[
  h_i = \tilde{h}_i + \tilde{f}\tilde{X}_{,i} + \tilde{g}\tilde{Y}_{,i}\ .
\]
For the function $R(U,V,o) = \tilde{R}(\tilde{U},\tilde{V},o)$, we obtain
\[
  R_{,U} = \tilde{R}_{,\tilde{U}}\tilde{X}_{,U}\ , \quad R_{,V} =
  \tilde{R}_{,\tilde{V}}\tilde{Y}_{,V}\ , 
\]
and 
\[
  R_{,i} = \tilde{R}_{,i} + \tilde{R}_{,\tilde{U}}\tilde{X}_{,i} +
  \tilde{R}_{,\tilde{V}}\tilde{Y}_{,i}\ . 
\]
The transformation of the logarithm is
\[
  \ln\left(-\frac{\tilde{U}'}{\tilde{V}'}\right) =
  \ln\left(-\frac{\tilde{X}_{,U}U'}{\tilde{Y}_{,V}V'}\right) =
  \ln\left(-\frac{U'}{V'}\right) +
  \ln\left(-\frac{\tilde{X}_{,U}}{\tilde{Y}_{,V}}\right) \ .
\]
Collecting all terms, we obtain
\begin{eqnarray}
  F & = & \frac{\partial}{\partial U} \left[\frac{R^2}{4}
  \ln\left(\frac{\tilde{X}_{,U}}{\tilde{Y}_{,V}}\right)\right]  
  - \frac{R^2}{4}\frac{\tilde{X}_{,UU}}{\tilde{X}_{,U}} +
  \tilde{F}\tilde{X}_{,U}\ , 
\label{8,2} \\
  G & = & \frac{\partial}{\partial V} \left[\frac{R^2}{4}
  \ln\left(\frac{\tilde{X}_{,U}}{\tilde{Y}_{,V}}\right)\right]  
  + \frac{R^2}{4}\frac{\tilde{Y}_{,VV}}{\tilde{Y}_{,V}} +
  \tilde{G}\tilde{Y}_{,V}\ , 
\label{9,1} \\ 
  H_i & = & \frac{\partial}{\partial o^i} \left[\frac{R^2}{4}
  \ln\left(\frac{\tilde{X}_{,U}}{\tilde{Y}_{,V}}\right)\right]  
  - \frac{R^2}{4}\left(\frac{\tilde{X}_{,Ui}}{\tilde{X}_{,U}} -
  \frac{\tilde{Y}_{,Vi}}{\tilde{Y}_{,V}}\right) + 
  \tilde{H}_i + \tilde{F}\tilde{X}_{,i} + \tilde{G}\tilde{Y}_{,i}\ . 
\label{9,2} 
\end{eqnarray}
The first terms on the right-hand sides of Eqs.~(\ref{8,2})--(\ref{9,2})
represent a divergence; according to the result of the previous section, these
terms can be thrown away, and we obtain finally:
\begin{eqnarray}
  F & = & \tilde{F}\tilde{X}_{,U} -
  \frac{R^2}{4}\frac{\tilde{X}_{,UU}}{\tilde{X}_{,U}}\ ,
\label{9,3} \\
  G & = & \tilde{G}\tilde{Y}_{,V} +
  \frac{R^2}{4}\frac{\tilde{Y}_{,VV}}{\tilde{Y}_{,V}}\ ,
\label{9,4} \\
  H_i & = & \tilde{H}_i + \tilde{F}\tilde{X}_{,i} + \tilde{G}\tilde{Y}_{,i} -
  \frac{R^2}{4}\left(\frac{\tilde{X}_{,Ui}}{\tilde{X}_{,U}} -
  \frac{\tilde{Y}_{,Vi}}{\tilde{Y}_{,V}}\right)\ .
\label{9,5} 
\end{eqnarray}
These equations yield the transformation of the functions $\tilde{F}$,
$\tilde{G}$ and $\tilde{H}_i$ that solve Eqs.~(\ref{6,5})--(\ref{6,7}) for the
gauge $\tilde{U},\tilde{V}$ to the functions $F$, $G$, and $H_i$ that solve
them for the gauge $U,V$.

\subsection{Integration of equations (\ref{6,5})--(\ref{6,7})}
The transformation equations (\ref{9,3})--(\ref{9,5}) can be used to calculate
the solutions $F$, $G$, and $H_i$ from some well-known solutions $\tilde{F}$,
$\tilde{G}$ and $\tilde{H}_i$. In fact, one can always choose the gauge
$\tilde{U},\tilde{V}$ in such a way that the right-hand sides of
Eqs.~(\ref{6,5})--(\ref{6,7}) simplify and become trivially solvable. Let us,
for instance, choose the gauge as follows.

Let the volume term to be transformed correspond to the flat spacetime. Then
we can choose the coordinates $\tilde{U}$ and $\tilde{V}$ as the retarded and
advanced time coordinates for Minkowski spacetime and
\[
  \tilde{A} = 1\ ,\quad \tilde{R} = \frac{-\tilde{U}+\tilde{V}}{2}\ .
\]
Eqs.~(\ref{6,5})--(\ref{6,7}) now read
\begin{eqnarray*}
  \tilde{F}_{,\tilde{V}} - \tilde{G}_{,\tilde{U}} & = & 0\ , \\
  \tilde{H}_{i,\tilde{U}} - \tilde{F}_{,i} & = & 0\ , \\
  \tilde{H}_{i,\tilde{V}} - \tilde{G}_{,i} & = & 0\ , 
\end{eqnarray*}
and we guess a solution to be
\begin{equation}
  \tilde{F} = \tilde{G} = \tilde{H}_{i} = 0
\label{11,1}
\end{equation}
for all $i$.

Let the volume term correspond to the $(\alpha\beta)$-quadrant (see I,
Sec.~2.1) of the Schwarzschild spacetime with the mass
parameter $M(o)$ (the mass parameter is Dirac's observable and so it is, in
general, a function of the chosen complete system of those observables). Then
we can choose $\tilde{U}^\alpha$ and $\tilde{V}^\beta$ to be the double-null
Eddington-Finkelstein coordinates defined by Eqs.~(1) and (2) of I so that
Eqs.~(3), (4) and (5) of I hold:
\begin{eqnarray}
  \tilde{A} & = & \left|1 - \frac{2M(o)}{\tilde{R}}\right| =
  \alpha\beta\left(1 - \frac{2M(o)}{\tilde{R}}\right)\ ,
\label{11,2a} \\
  \tilde{R} & = & 2M(o)\kappa
  \left[\alpha\beta\exp\left(\frac{-\alpha\tilde{U}^\alpha + 
  \beta\tilde{V}^\beta}{4M(o)}\right)\right]\ ,
\label{11,2b}
\end{eqnarray}
where $\kappa$ is defined by Eq.~(5) of Ref.~\cite{I}. On substituting
Eqs.~(\ref{11,2a}) and (\ref{11,2b}) for $A$ and $R$ into 
Eqs.~(\ref{6,5})--(\ref{6,7}), we obtain
\begin{eqnarray*}
  \tilde{F}_{,\tilde{V}} - \tilde{G}_{,\tilde{U}} & = & 0\ , \\
  \tilde{H}_{i,\tilde{U}} - \tilde{F}_{,i} & = & -\frac{\alpha}{2}M_{,i}\ , \\
  \tilde{H}_{i,\tilde{V}} - \tilde{G}_{,i} & = & -\frac{\beta}{2}M_{,i}\ .
\end{eqnarray*}
Again, we guess easily
\begin{eqnarray}
  \tilde{F} & = & 0\ , \\
\label{11,3a}
  \tilde{G} & = & 0\ , \\
\label{11,3b}  \\
  \tilde{H}_i & = & -\frac{\alpha\tilde{U}^\alpha +
  \beta\tilde{V}^\beta}{2}M_{,i}\ . 
\label{11,4}
\end{eqnarray}

The double-null Eddington-Finkelstein gauge may be simple, but it is singular
($\tilde{U}^\alpha = \alpha\infty$ at the future and $\tilde{V}^\beta =
-\beta\infty$ at the past horizon). The solution (\ref{11,3a}), (\ref{11,3b})
and (\ref{11,4}) diverges at both horizons. This singularity can, however, be
removed by subtracting a suitable $W$-term. To show this, let us transform to
a regular gauge, for example to the Kruskal coordinates $U$ and $V$:
\[
  U = -\exp\left(-\frac{\alpha\tilde{U}^\alpha}{4M}\right)\ ,\quad V =
  \exp\left(\frac{\beta\tilde{V}^\beta}{4M}\right)\ . 
\]
Then
\begin{eqnarray*}
  \tilde{X}(U,o) & = & -4\alpha M(o)\ln(-U)\ , \\
  \tilde{Y}(U,o) & = & 4\beta M(o)\ln(V)\ .
\end{eqnarray*}
An easy calculation using Eqs.~(\ref{9,3})--(\ref{9,5}) yields
\[
  F = \frac{R^2}{4U}\ ,\quad G = -\frac{R^2}{4V}\ ,\quad H_i =
  -2MM_{,i}\ln\left(\frac{-U}{V}\right)\ .
\]
These functions are singular at the horizons ($U=0$ or $V=0$). However, if we
choose
\[
  W(U,V,o) = -M^2(o)\ln\left(\frac{-U}{V}\right)\ ,
\]
then the equivalent solution defined by
\[
  F_{\text{reg}} := F + W_{,U}\ ,\quad G_{\text{reg}} := G + W_{,V}\ ,\quad 
  H_{i\text{reg}} := H_i + W_{,i}
\]
is regular everywhere. We obtain (using some properties of the function
$\kappa$, see Eq.~(12) and (51) of \cite{H-Kie})
\begin{eqnarray}
  F_{\text{reg}} & = & - \frac{M(R+2M)}{2\exp(R/2M)}V\ , 
\label{16,1} \\
  G_{\text{reg}} & = & \frac{M(R+2M)}{2\exp(R/2M)}U\ , 
\label{16,2} \\
  H_{i\text{reg}} & = & 0\ ,
\label{16,3}
\end{eqnarray}
where
\[
  R = 2M\kappa(-UV)\ .
\]

The solutions (\ref{11,1}) and (\ref{11,3a})--(\ref{11,4}) or
(\ref{16,1})--(\ref{16,3}) together with the formulae (\ref{9,3})--(\ref{9,5})
can help us to calculate the functions $F$, $G$ and $H_i$ for the two-shell
system. For the single shell, the resulting formulae of \cite{H-Kie} can
easily be reproduced by the new method.

\section*{Acknowledgments}
The authors are thankful for useful discussions by C.~Kiefer and K.~V.~
Ku\-cha\v{r}. The work has been supported by the Swiss Nationalfonds and by
the Tomalla Foundation, Zurich.

\end{document}